# Precise Gravitational Tests via the SEE Mission: A Proposal for Space-Based Measurements

A. J. Sanders
Department of Physics and Astronomy
University of Tennessee
Knoxville, TN 37996
asanders@utk.edu

G. T. Gillies
School of Engineering and Applied Science
University of Virginia
Charlottesville, VA 22901
gtg@virginia.edu

ABSTRACT

The objective of a SEE mission is to support development of unification theory by carrying out sensitive gravitational tests capable of determining whether various alternative theories are compatible with nature. Gravitation is a key "missing link" in unification theory. Nearly all unification theories incorporate gravity at a fundamental level, and therefore precise measurements of gravitational forces will place important constraints on unification theories. Ground-based gravitational measurements to the accuracy required are impossible due to the many sources of noise present in the terrestrial environment. The proposed space-based Satellite Energy Exchange (SEE) mission will measure several important parameters to an accuracy between 100 and 10,000 times better than current or planned measurement capabilities. It will test for time variation of the gravitational "constant" $G$ and for violations of the weak equivalence principle (WEP) and the inverse-square-law (ISL), and it will determine $G$. It is well-known that the discovery of breakdowns in WEP or ISL and the possible determination of a time-varying G would have significant consequences on virtually all aspects of unification theory.

1. Introduction

SEE is an international effort to plan and conduct a next-generation space mission designed to measure weak gravitational effects and to make fundamental tests of theories. SEE will search for a time variation of the gravitational constant $G$, test for inverse-square-law (ISL) violations (at ranges of $\sim 1$ and $\sim 10^6$ m), and determine the gravitational constant $G$, all at target accuracies that are aimed at reducing present experimental uncertainties by factors of $10^2$ to $10^4$. Analysis of data from the SEE mission will also automatically yield tests of the Weak Equivalence Principle (WEP) at both ranges, based on possible composition dependence (CD), but without a the same substantial decreases in uncertainty.

The SEE mission will entail launching a free-flying observatory, the heart of which will be an experimental test chamber in which two or three test bodies (one large "Shepherd" and one or two small "Particles," depending on the experiment) will float freely, experiencing only each other's gravity and that of the Earth and other bodies in the solar system. Observing perturbations of the test bodies will provide the data required to achieve the scientific objectives of the SEE mission.

The focus of SEE is post-Einsteinian: it is not limited to specific tests of general relativity, thus making it a type of *next-generation mission,* in that it will measure or test a number of links of gravitation important to the development of unified theories. As such, the SEE mission has six measurement goals. These goals and the target accuracies for the measurements are summarized in Table 1.

*Table 1: Expected Accuracy of SEE Tests and Measurements*

| Test/Measurement | Expected Accuracy | Comment |
|---|---|---|
| (G-dot)/G | ~$10^{-14}$/yr in one year | 100 times better than current bound; likely to discriminate against most modern theories |
| ISL at ~few meters | $2 \times 10^{-7}$ | 1,000 to 10,000 times better than current bound |
| ISL at ~$R_E$ | $1 \times 10^{-10}$ | 100 times better than current bound |
| G | 0.33 ppm (330 ppb) | 100 times more precise in single day than ground-based experiment in months |
| WEP at ~few meters | $<10^{-7}$ ($\therefore \alpha<10^{-4}$) | Comparable to ground-based experiments |
| WEP at ~$R_E$ | $<10^{-16}$ ($\therefore \alpha<10^{-13}$) | 1,000 to 10,000 better than current bound but 100 times weaker than STEP |

Although our long-range WEP test will be at a somewhat lower accuracy than those expected from STEP and Mircroscope, it is nevertheless designed to achieve three to four orders better than existing tests. Moreover, if either STEP or Mircroscope finds a large violation, the need for confirmation will be urgent, and the SEE test of WEP could play a useful role in achieving it.

2. Proposed ISL and WEP Tests

2.1 Background

Tests of *both* the inverse-square-law (ISL) and of the weak equivalence principle (WEP) by composition-dependence (CD) are extremely important because of the far-reaching and profound consequences of any violation. Any apparent violation could be readily interpreted as evidence of the existence of a new super-weak force, presumably of short range and therefore mediated by a massive particle—*i.e.*, in terms of a Yukawa-type potential:

$$U(r) = (GM/r) \times [1 - \alpha \exp(-r/\Lambda)] \tag{1}$$

where α is the interaction strength and Λ is the range. However, if the experimental data from some mission indicate an apparent violation but do not fit the Yukawa hypothesis, then other alternatives must of course be considered.

The Yukawa model (Eq. 1) has become the standard and customary way to parameterize possible apparent WEP violations. This approach unites both ISL and CD effects very naturally, while the parameter values in the Yukawa potential suggest what experimental conditions will permit sensitive tests of the ISL and WEP. Nevertheless, we must stress that it is especially important to do ISL tests, *even for values of Λ where CD tests have already set tight limits on α,* because the model-dependence of ISL tests is different from that of CD tests (Sanders & Deeds, 1992a, 1992b).

2.2  Terrestrial Tests of ISL and WEP

In the watershed year of 1986, Fischbach startled the physics community by showing that Eötvös' famous turn-of-the-century experiment is much less decisive as a null result than was generally believed (Fischbach *et al.*, 1986). Prior to this time, experiments by Dicke (Roll *et al.*, 1964) and Braginsky (1977) had tested the universality of free fall (UFF) to very high accuracy with respect to several metals falling in the gravitational field of the sun. The interpretation of these results at the time was that they validated UFF: It was implicit that any violation would have infinite range, like gravity (Adelberger, 1994). During the 1970s and early 1980s there had also been a flurry of activity concerning possible ISL violations, which eventually led to null results at the levels of precision then available (Fujii, 1971 and 1972; Long, 1976 and 1984).

Following Fischbach's 1986 conjecture, many investigators undertook searches for either ISL or WEP violations. Although a number of anomalies were initially reported, nearly all of these were eventually explained in terms of overlooked systematic errors or extreme sensitivity to models, while most investigators obtained null results (Fischbach & Talmadge, 1999).

Only one experiment—a WEP experiment by Thieberger (1987)—has not succumbed. However, this result is widely presumed to need further investigation, probably in part because a null result was subsequently obtained by a similar experiment (Bizzeti *et al.*, 1989; Bizzeti *et al.*, 1990) that was designed to be more sensitive (albeit on the basis of a rather literal faith in current fifth-force models), and in part because it is difficult to understand how this one result could stand in contrast to so many other experiments that found no non-Newtonian effect, even with bounds that are tighter (in the context of these models) than those of Thieberger.

By far the tightest laboratory bounds are those obtained by Adelberger and his "Eöt-Wash" group at the University of Washington. Their recent limit on ISL violations at very short distances is a factor of 1,000 improvement over previous results (Hoyle et al., 2001); their limit on WEP is presently ~$10^{-13}$ (Adelberger *et al.* 1987 and 1990; Adelberger 1994; Su *et al.*, 1994; Smith et al., 2000; Adelberger, 2001; and Adelberger et al., 2003). Their bound on the interaction strength $\alpha$ is approaching 1 part in $10^{-9}$ for the long-range ($\Lambda > R_E$) WEP test (Adelberger, 1997). This result, like all WEP tests, is to some extent model-dependent. For example, the limits on $\alpha$ very often are based on the assumption that the mediator is a *vector* boson and that the mixing angle $\theta_5$ is zero (*i.e.*, the fifth force "charge" is simply the boson number: $\theta_5 = B$), and this is what is most commonly shown in Gibbons-Whiting diagrams.

We note that Lunar Laser Ranging (LLR) has previously validated UFF, using the Nordtvedt effect, to $5 \times 10^{-13}$ (Williams *et al.*, 1996; Dickey *et al.*, 1994) and now $1.5 \times 10^{-13}$ (Anderson & Williams, 2001), and that another experiment at U. Wash. to test the ISL is now ongoing (Boynton, 2000). Nevertheless, a rather paradoxical situation has developed: The faith of gravitational physicists in the WEP and ISL has weakened since 1986, despite not only the dearth of contrary experimental evidence, but also the actual tightening of the bounds on possible violations. This conceptual shift can reasonably be attributed to the influence of theory development during this period: There is a growing belief, now approaching a consensus, that unification theories (supergravity, superstrings, M-theory, etc.) are likely to require some form of violation of the inverse-square law and/or the equivalence principle. Thus, the search for violations continues. For reviews of terrestrial searches for non-Newtonian gravity, see Fischbach, Gillies, *et al.* (1992), Franklin (1993), Adelberger (1994 and 2001), and Fischbach & Talmadge (1999).

3. SEE Methodology

3.1 Technical Summary

The SEE mission will measure the orbital perturbations of a large test body (the "Shepherd") and one or more small test bodies ("Particles") in a near-zero-g environment. This is accomplished by placing the test bodies within a spacecraft, which shields them from radiation pressure while providing a drag-free, uniform-temperature environment for the experiments. The positions of the test bodies are tracked in the coordinate frame of the experimental chamber (see Figure 1), while the position of the spacecraft is tracked in the frame of the Earth by GPS or other comparable ground tracking methods, such as SLR. The forces on the test bodies are inferred from their orbital perturbations. The Particle(s) is(are) launched, retrieved, and re-launched approximately once per day. The Shepherd remains in orbit continuously, and the SEE observatory is "slaved" to the Shepherd, although it is highly decoupled from the observatory.

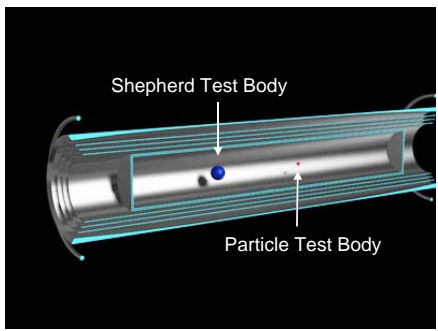

Figure 1: A cross section of the SEE test chamber showing the "Shepherd" and "Particle" masses.

3.2 Dynamics of the SEE Trajectories

The relative motion of the two test bodies--the large "Shepherd" and the small "Particle"--during a SEE encounter can appear rather paradoxical. Two co-orbiting bodies--one trailing the other--may exchange substantial gravitational energy if their orbits are nearly identical, so that they remain very close together for several orbits around the Earth. If the body in the lower (and therefore faster) orbit approaches the other from behind, the trailing body is picking up energy from the leading body and, after the passage of some time, may acquire sufficient energy to rise above the leading body (stated more precisely, the semi-major axis of the orbit of the trailing body may grow to exceed that of the leading body). At this point the trailing body will begin to *fall back*, while still continuing to pick up energy from the leading body. Thus, although the gravitational force is of course always attractive--a SEE encounter gives the paradoxical appearance of mutual *repulsion* by the two bodies. This phenomenon can be understood in terms of the virial theorem (Sanders & Deeds, 1992a). It was first discovered by Darwin a century ago (Darwin, 1897).

The experimental masses and distances must be scaled so that the duration of a SEE encounter is long compared to the orbital period (e.g., about one day), since the encounter must be adiabatic. Choosing the Shepherd mass to be ~100 to 250 kg makes the change in the altitude $\Delta r$ of the small test body (the Particle) during an encounter typically a small multiple of 10 cm, which is convenient. We find that $\Delta r$ is proportional to the mass of the Shepherd and approximately inversely proportional to the distance of closest approach. If the permitted encounter length is ~5 to 10 meters, then the encounter duration is ~1 day, and the instantaneous relative speeds are typically 100 to 300 $\mu m\ s^{-1}$. Note that a SEE-encounter trajectory is rather long and narrow--in fact, it is virtually one-dimensional. The extremely narrow shape of a SEE trajectory has an important consequence: nearly all the information is contained in the *separation* of the test bodies as a function of time. This fact greatly simplifies data analysis, and results in relaxed observatory pointing requirements (Sanders et al., 1999).

3.3 SEE Spacecraft Requirements

The SEE spacecraft must be designed to protect the test bodies from all forces except gravity--each other's gravity and that of the Earth and other bodies in the solar system--and to provide a method for accurate position determination of the test bodies. This requires that the spacecraft provide:

1) A drag-free environment for the test bodies during their encounters,
2) A long, cylindrical internal volume for the test body trajectories,
3) A nearly zero-g internal gravitational field,
4) A uniform thermal environment for the test bodies,
5) Shielding to minimize charging of the test bodies,
6) A source of electrons to neutralize the charge,
7) Support for the precision test body location measurement systems, and
8) Support for robotic launch and retrieval of the test bodies.

Magnetic shielding of the test bodies may also be required, although we currently do not expect magnetic perturbations to be significant. Although challenging, the spacecraft requirements appear feasible without the need for new technology developments. One of the objectives of the proposed study will be to better define and flow-down the spacecraft requirements in more detail, perform trade studies of potential implementation approaches, and develop a preliminary spacecraft concept that satisfies the SEE mission requirements.

4. Testing ISL and WEP via the SEE Interaction

4.1 Overview of the Approach

The SEE mission will entail precise analyses of the *relative* motion of the two test bodies during a number of SEE encounters. We expect that, with some exceptions, the analysis of these observations can be performed almost independently of ground tracking. The value of *G* is obtained from the accelerations of a Particle relative to the Shepherd during a SEE encounter. Thus, account is taken of the peculiar and counter-intuitive dynamics which results from the fact that both bodies are in orbit around the *Earth* and are chiefly under the influence of its gravity rather than each other's. The intermediate-range (~meters) inverse-square-law (ISL) test will straightforwardly compare the measured values of *MG* (*M* is the Shepherd mass) obtained at various locations along the trajectory of each SEE encounter, and then search for apparent variation of *MG* as a function of the separation of the test bodies. We note that uncertainties in the Shepherd mass *M* drop out. The long-range (~radius of the Earth) ISL test takes advantage of the fact that perigee precession would be caused by the perturbing force of a putative Yukawa-type particle. This precession is due mainly to the cubic term in the force, in exact analogy to the anomalous precession of Mercury predicted by general relativity.

We know of two distinct ways of detecting such a precession: The original method (Sanders and Deeds, 1992a) involves observing a *relative* precession of the perigees of the test bodies by detailed analysis of the SEE-encounter trajectory, which is based on the measurements of the relative positions of the test bodies within the experimental test chamber. The "signal" in this case is a slight difference between the apsidal and sidereal periods of the Particle mass. More recently Nordtvedt (1998) has suggested an improved method based upon ground tracking specific to the SEE mission. The SEE WEP tests based on composition difference are not expected to be at path-breaking accuracy at either the intermediate (~ meters) or long range (~ radius of the Earth). The methods were described in Sanders & Deeds (1992a). We have subsequently realized that the long-range test should be altered in two respects, viz. by comparing two different Particles in flight simultaneously rather than by comparing the Shepherd with a Particle, and by briefly using continuous interferometry in order to obtain their differential radii far more accurately than our micron-ranged tracking could do.

4.2 Example of an Assessment of a Technical Issue: Thermal Limits

There are clearly many fundamental problems that must be addressed in the design of a space-based measurement of this type. We present only one example of an interesting such issue that will underlie the potential success of the SEE approach: thermal modeling of the SEE observatory by a series of proxies of increasing complexity. A key step in undertaking this work was to determine that the thermal conductivity in the walls of the experimental chamber will spread the heat axially well enough to achieve our uniformity goal of 1 mK. This work was done in cooperation with the NASA Marshall Space Flight Center. Table 2 summarizes the modeling by NASA MSFC colleagues of the temperature on the SEE experimental chamber walls. Reading along any row, we see that the uniformity is ~100 µK, which is better than required. Other key findings shown by the table: The chamber does reach essentially liquid nitrogen—by purely passive means. However, the temperature variation as a function of the sun angle β indicates that a small amount of active heating or cooling will be required to prevent oscillation of temperature with position in orbit. Subsequent studies by the SEE team considered explicitly the effects of observatory rotation and the magnitude of conductive (non-radiative) heat transfers between the successive concentric cylinders along the support structures (presumed to be either plastic struts with thermal conductivity twice that of asbestos or titanium bicycle spokes). These studies validated the results of prior studies, which had used simpler models of the observatory.

**Table 3: Axial Temperature Distribution of Experimental-Chamber Walls in SEE Satellite**

Thermal Conductivity = 150 W/mK, Solar Flux = 1353 W/m$^2$, $\alpha$ = 0.14, $\varepsilon$ = 0.90

| Station No.* | 1 | 2 | 3 | 4 | 5 | 6 | 7 | 8 | $T_{max} - T_{min}$ |
|---|---|---|---|---|---|---|---|---|---|
| $\beta = 90°$ | 81.5581 | 81.5580 | 81.5580 | 81.5580 | 81.5580 | 81.5580 | 81.5580 | 81.5580 | 0.2 mK |
| 85° | 81.5573 | 81.5573 | 81.5572 | 81.5572 | 81.5572 | 81.5572 | 81.5572 | 81.5573 | 0.1 |
| 80° | 81.5580 | 81.5580 | 81.5579 | 81.5579 | 81.5579 | 81.5579 | 81.5579 | 81.5579 | 0.1 |
| 75° | 81.5571 | 81.5570 | 81.5569 | 81.5569 | 81.5569 | 81.5569 | 81.5569 | 81.5569 | 0.2 |
| 70° | 81.5512 | 81.5511 | 81.5510 | 81.5510 | 81.5509 | 81.5509 | 81.5509 | 81.5509 | 0.3 |
| 65° | 81.3505 | 81.3504 | 81.3503 | 81.3503 | 81.3502 | 81.3502 | 81.3502 | 81.3502 | 0.3 |
| 60° | 80.9458 | 80.9457 | 80.9457 | 80.9456 | 80.9455 | 80.9455 | 80.9455 | 80.9455 | 0.3 |
| 55° | 80.5248 | 80.5248 | 80.5347 | 80.5246 | 80.5245 | 80.5245 | 80.5245 | 80.5245 | 0.3 |

* Stations are evenly spaced along *z*-axis, with station 1 nearest the Sun.

4.3  Possible Cosmological Implications

Moffat and Gillies (2002) have suggested the possibility SEE be used to explore whether or not the cosmological constant arises from zero point energy. This would be done via a scenario in which the violation of the weak equivalence principle (WEP) that would result from such an effect might be observed via the SEE interaction as discussed above. According to this proposal, the acceleration of a spherical test mass of aluminum would be compared with that of a similar test mass made from another material (e.g., a monel metal like copper or silver). Aluminum is chosen because, as observed by others, it has a relatively sharp transition from reflectance to absorption of electromagnetic waves at photon energies of approximately 15.5 eV. At this energy, the ratio of the zero-point energy density inside the aluminum sphere to the rest mass energy density of it is ≈ 1.6 x $10^{-14}$. Therefore, if comparisons were made between test masses of aluminum and, e.g., copper, a violation of WEP at approximately this level should be observed if the zero point energy does not couple to the gravitational field. Such an experiment would be a direct test of the role that a purely quantum mechanical effect plays in general relativity.

5. Discussion and Conclusions

Most of the current promising approaches to unification theory, including string theories, p-brane theories, and supergravity, contain gravity at a fundamental level. Although a number of different theoretical schemes have been proposed, a lack of precise experimental evidence presently makes it almost impossible to assess the validity of alternative schemes. Gravitation is the missing link in efforts to achieve a satisfactory unification theory of physics. The very precise experimental data hoped for in the SEE mission might then augur for major advances in unification theory.

The SEE mission results will either:
► give extremely accurate confirmation of presently-accepted theories, or
► indicate violations of them, while suggesting the direction of necessary changes.

The central aspiration of the work is that precise new data that might be provided by the SEE mission will expose conflicts with some existing theories, thus revealing which theories are consistent with both the new evidence and previous evidence. This process is vital for identifying the most promising directions for further developments in unified theories.

Acknowledgments

This work was supported in part by grants from the U.S. National Aeronautics and Space Administration Fundamental Physics in Microgravity Program, NASA Marshall Space Flight Center, and NATO Division des Afffaires scientifiques et de l'Environnement, ESTEC. We are pleased to acknowledge continuing interest by NASA Jet Propulsion Laboratory. We thank S. W. Allison and M. R. Cates of Oak Ridge National Laboratory for several useful discussions.References